\documentclass[apjl,a4paper,12pt,useAMS]{emulateapj}
\usepackage{multirow}
\usepackage{booktabs}
\usepackage{makecell}
\usepackage{natbib}
\usepackage{array}
\newcommand{\tabincell}[2]{\begin{tabular}{@{}#1@{}}#2\end{tabular}}

\def\cm{\,\rm cm}
\def\s{\,{\rm s}}

\def\mpc{\,{\rm Mpc}}
\def\msun{\,{\rm M_{\odot}}}
\def\erg{\,{\rm erg}}
\def\kev{\,{\rm keV}}
\def\mev{\,{\rm MeV}}

\def\yr{\,{\rm  yr}}

\def\mpc{\,{\rm Mpc}}
\def\gpc{\,{\rm Gpc}}

\def\aj{Astronomical Journal }    
\def\apj{Astrophysical Journal }  
\def\apjl{Astrophysical Journal } 

\def\apjs{ApJS }                  
\def\mnras{MNRAS }                
\def\prd{Phys.~Rev.~D }           
\def\prl{Phys.~Rev.~Lett }        
\def\araa{ARA\&A }                
\def\nat{Nature }                 
\def\aap{A\&A }                   

\def\jcap{JCAP}
\begin{document}
\title{The jet structure and the intrinsic luminosity function of short gamma-ray bursts}
\author{Wei-Wei Tan$^{1,2}$ \& Yun-Wei Yu$^3$}

\altaffiltext{1}{School of Physics and Mechanical Electrical \& Engineering, Hubei University of Education, Wuhan 430205, China}
\altaffiltext{2}{Research Center for Astronomy, Hubei University of Education,
Wuhan 430205, China}
\altaffiltext{3}{Institute of Astrophysics, Central China Normal
University, Wuhan 430079, China, {yuyw@mail.ccnu.edu.cn}}

\begin{abstract}
The joint observation of GW170817 and GRB 170817A indicated that short gamma-ray bursts (SGRBs) can originate from binary neutron star mergers. Moreover, some SGRBs could be detected off-axis, while the SGRB jets are highly structured.
Then, by assuming an universal angular distribution of the jet emission for all SGRBs, we re-produce the flux and redshift distributions of the cosmological SGRBs detected by {\it Swift} and {\it Fermi}. For self-consistency, this angular distribution is simultaneously constrained by the luminosity and event rate of GRB 170817A. As a result, it is found that the universal jet structure of SGRBs could approximately have a two-Gaussian profile. Meanwhile, the intrinsic luminosity function (LF) of the on-axis emission of the jets can be simply described by a single power law with a low-luminosity exponential cutoff. The usually discovered broken-power-law apparent LF for relatively high luminosities can naturally result from the coupling of the intrinsic LF with the angular distribution of the jet emission, as the viewing angles to the SGRBs are arbitrarily distributed.
\end{abstract}


\section{Introduction}

The joint observation of the gravitational wave event GW170817 and the short gamma-ray burst (SGRB) GRB 170817A \citep{Abbott2017a,Goldstein2017} had confirmed the long-standing hypothesis that SGRBs are produced by mergers of double neutron stars and, also possibly, of neutron star-black hole binaries. This hypothesis was firstly suggested about three decades ago \citep{Paczynski1986,Eichler1989, Narayan1992}.

However, somewhat abnormally, the prompt luminosity of GRB 170817A of $\sim10^{47}\rm erg~s^{-1}$ is actually hundreds of times lower than the lowest luminosity of the normal cosmological (i.e., $z\gtrsim0.1$) SGRBs \citep[][]{Abbott2017b,Zhang2018}. This promoted some people to consider that GRB 170817A could belong to a new SGRB population \citep[e.g., ][]{Rueda2018}. As one possibility, it was suggested that GRB 170817A could be produced by a mildly-relativistic, radially structured, and isotropic outflow \citep[e.g., a cocoon powered by a chocked jet; ][]{Kasliwal2017,Mooley2018,Nakar2018}, rather than by a traditional relativistic jet \citep[e.g, ][]{Paczynski1986,Eichler1989,Meszaros1992,Narayan1992}. Nevertheless, in fact, the existence of a relativistic jet had been strongly favored by the observed apparent superluminal motion of the radio counterpart, which was discovered with the Very Long Baseline Interferometry \citep{Mooley2018,Ghirlanda2019}. Therefore, a more plausible and acceptable scenario is that the GRB outflow is highly angularly structured, e.g., consists of a relativistic jet core and a mildly-relativistic jet wing \citep[][]{Zhang2002,Kathirgamaraju2018,Resmi2018}. In this case, the observed GRB emission is sensitive to the viewing angle $\theta_{\rm v}$ relative to the outflow symmetric axis. The larger the viewing angle, the lower the GRB luminosity. Therefore, the low luminosity of GRB 170817A is just due to an off-axis observation. If it can be observed on-axis, then it will not be intrinsically different from the cosmological SGRBs. In other word, GRB 170817A is considered to have an origin identical to that of the cosmological SGRBs and its jet structure could be universal for all SGRBs.

The off-axis structured jet model for GRB 170817A had also been strongly supported by the temporal behavior of its multi-wavelength afterglows, which started to be detected from 9 days after the GW event \citep{Haggard2017,Hallinan2017,Margutti2017,Lyman2018,Mooley2018,Ruan2018,Troja2018}. The afterglow emission increased steadily until a peak at about 150 days and then turned to decrease \citep{Gill2018,Lazzati2018,Mooley2018}. Theoretically, afterglow emission can be detected only if its radiation cone can intersect the line of sight, which requires its Lorentz factor has decreased from the initial value to $\Gamma\sim \theta_{\rm v}^{-1}$ for an off-axis observer. Therefore, the brighter emission from the faster material closer to the jet center would appear later, which leads to the steady growth of the flux until the center of the jet is detected. Then, by fitting the afterglow light curves, the angular distributions of the energy and the Lorentz factor of the jet had been well constrained, which disfavored the isotropic and the top-hat jet models completely \citep{Kathirgamaraju2017,
Lamb2017, Alexander2018, DAvanzo2018, Gill2018, Beniamini2019, Howell2019, Lazzati2018, Lyman2018, Margutti2018, Resmi2018, Troja2018, Xie2019, Kathirgamaraju2019, Wu2019}. As the most important parameters, the opening angle of the jet of GRB 170817A was constrained to be around $\theta_{\rm j}\sim2^\circ-9^\circ$, which may have different definitions in different works, and the viewing angle is about $\theta_{\rm v}\sim20^\circ-30^\circ$ \citep[][]{Lamb2017,Lazzati2018,Lyman2018,Margutti2018,Resmi2018,Troja2018,Ghirlanda2019}. The latter one is well consistent with the constraint from the GW analysis \citep{Abbott2017a}.

The fact that the viewing angle is much larger than the jet opening angle can somewhat help to understand the high event rate of $\dot{R}_{\rm GRB 170817A}\sim190^{+440}_{-160}\rm yr^{-1}Gpc^{-3}$ of this unusual SGRB \citep{Zhang2018}. Obviously, this rate is significantly higher than the previous estimates of the local SGRB rate $\dot{R}_{\rm SGRB}(0)$ ranging from a few to a few ten $\rm yr^{-1}Gpc^{-3}$, which were obtained from the statistics of the cosmological SGRBs ($z\gtrsim 0.1$) for an adopted minimum luminosity at $\sim 10^{49}\erg \s^{-1}$  \citep{Guetta2006,Nakar2006,Guetta2009,Dietz2011,Coward2012,Wanderman2015,Tan2018,ZhangGQ2018}. As a straightforward impression, the difference between the traditional $\dot{R}_{\rm SGRB}(0)$ and $\dot{R}_{\rm GRB170817A}$ could be very roughly explained by the ratio of $(1-\cos \theta_{\rm j})/(1-\cos\theta_{\rm v})$, which can be true if the cosmological SGRBs are all on-axis and the on-axis luminosity of GRB 170817A is just the most probable luminosity of SGRBs. However, as inferred from the afterglow fittings, the on-axis luminosity of GRB 170817A is probably as high as $\sim10^{52}\rm erg~s^{-1}$ \citep[e.g., ][]{Lazzati2018, Lyman2018, Margutti2018, Resmi2018, Troja2018, Howell2019}, which is much higher than the luminosities of a large number of cosmological SGRBs. Therefore, if all SGRBs including GRB 170817A indeed have a common origin, then it can be concluded that a remarkable fraction of the cosmological SGRBs were actually observed off-axis for different viewing angles. In other word, the observational luminosity distribution of SGRBs, which can be substantially influenced by the angular distribution of the jet emission, must be significantly different from the intrinsic luminosity function (LF) of SGRBs. In this case, the event rates of the cosmological SGRBs also need to be re-estimated.

Therefore, the purpose of this paper is to combine GRB 170817A and the cosmological SGRBs into a united model, by invoking an universal angular distribution of the jet emission of all SGRBs. Specifically, by incorporating with the implications of GRB 170817A for the jet structure and the local event rate, we revisit the modeling of the flux and redshift distributions of the cosmological SGRBs detected by the {\it Swift} and {\it Fermi} observatories, so that the intrinsic LF is further constrained. These combined constraints can deepen our understanding of the jet formation and enable us to forecast more accurately the observational prospects of some future facilities in the opening GW astronomy era.

\section{The basic model assumptions}
The structure of a GRB jet is determined by both the jet launching mechanism and the propagation of the jet through the ambient material \citep{Kathirgamaraju2017, Lamb2017,
Beniamini2019, DAvanzo2018, Lazzati2018, Margutti2018, Xie2019}. For long GRBs, such an ambient material specifically refers to the envelope of the progenitor star and, sometimes, plus the stellar wind. For SGRBs, the relevant ambient material is ejected in about 0.1 second by the merging neutron stars through several channels including the tidal centrifugation, the collision squeeze, and the accretion disk wind \citep{Rosswog2005,Oechslin2006,Kasen2017}. The interaction of a GRB jet with a pre-existing ejecta can lead to a forward shock sweeping up the ejecta material. Then the swept-up ejecta material can flow out laterally to surround the jet, which is usually named as a cocoon. The high-pressure of the cocoon could collimate the jet effectively and the degree of the collimation depends on the ratio of the jet luminosity to the column density of the ambient material \cite[e.g.,][]{Yu2020}. Sometimes, the jet can be choked finally by the ambient material, if the central engine of the SGRB is switched off before the jet head reaches the surface of the merger ejecta. In this case, all of the jet energy is absorbed by the cocoon. Then, due to its high temperature, the cocoon could still emerge from the merger ejecta to form a wide-angle and mildly relativistic flow. Such a cocoon flow has been frequently suggested to account for the prompt and afterglow emissions of GRB 170817A, if its energy can have a fine-tuned radial distribution \citep{Kasliwal2017, Mooley2018, Nakar2018}. Nevertheless, as discussed in the introduction, it is more probable that the jet of GRB 170817A had broken out from the merger ejecta successfully. After the jet breakout, both the jet and the cocoon can expand freely in the low-density interstellar medium. The sharp drop of the cocoon pressure relieves the jet from the collimation. Then, the opening angle of the jet can increase quickly, until it reaches the initial value or until the jet injection is stopped.

\begin{figure}
 \centering\resizebox{0.45\textwidth}{!}
	{\includegraphics{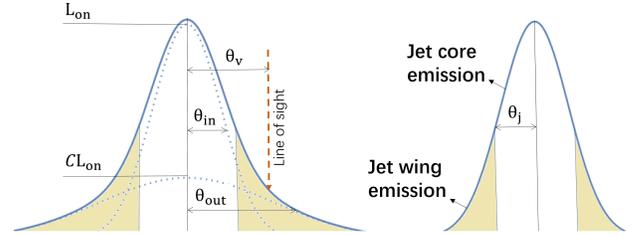}}
    \caption{An illustration (not in scale) of the two-Gaussian angular distribution of the isotropically-equivalent luminosity of a SGRB jet (left), in comparison with a traditional single-Gaussian distribution (right).}
    \label{fig1}
\end{figure}

As a result of the jet-cocoon interaction, it is natural to expect that a breakout GRB jet can consist of a relativistic beamed core and a mildly-relativistic widely-spreading wing. It could be not easy and actually unnecessary to separate the cocoon component from the jet and, thus, we might as well include the cocoon contribution into the wing of the jet. Therefore, the structure of a breakout jet can be sometimes simplified to a two-component jet \citep{Frail2000, Berger2003,
Racusin2008, Filgas2011}. Or, for a continuous description, the angular distributions of the energy and the Lorentz factor of the GRB jets were also suggested to have a power-law decaying \citep{Dai2001, Zhang2002, Kumar2003} or a Gaussian function profile \citep{Zhang2002, Kumar2003, Rossi2004}. In this paper, we assume that the isotropically-equivalent luminosity of SGRBs observed at different viewing angles ($\theta_{\rm v}$) can be described by an empirical function including an inner and an outer Gaussian component as:
\begin{eqnarray}
L_{\rm iso}(\theta_{\rm v})=L_{\rm on}\left[\exp\left(-{\theta_{\rm v}^2\over 2\theta_{\rm in}^2}\right)+\mathcal C\exp\left(-{\theta_{\rm v}^2\over 2\theta_{\rm out}^2}\right)\right],\nonumber\\ \label{LDIS_total}
\end{eqnarray}
which is determined by four free parameters $L_{\rm on}$, $\theta_{\rm in}$, $\theta_{\rm out}$, and $\mathcal C$ as labeled in Figure \ref{fig1}. This expression could roughly reflect the general structure of SGRB jets exhibited in the hydrodynamical simulations \citep[e.g., ][]{Lazzati2017, Salafia2020}. However, please keep in mind that the $L_{\rm iso}(\theta_{\rm v})$ function, which can be compared with observations directly, could be somewhat different from the intrinsic angular distribution of the jet energy $E(\theta)$,\footnote{By assuming the GRB duration $T_{\rm GRB}$ to be independent of the observational direction, the isotropically-equivalent luminosity for the observer at $\theta_{\rm v}$ can be calculated by
\begin{eqnarray}
L_{\rm iso}(\theta_{\rm v})=4\pi\int {dL'\over d\Omega'}{1\over \Gamma^4(\theta)[1-\beta(\theta)\cos\alpha]^4}\sin\theta d\theta d\phi,\nonumber
\end{eqnarray}
where the spherical coordinates $(\theta,\phi)$ are defined relative to the jet symmetric axis. $dL'/d\Omega'=\eta_{\gamma}E(\theta)/(4\pi T_{\rm GRB})$ is the radiation intensity in the comoving frame with $\eta_{\gamma}$ being the radiation efficiency. The angle $\alpha$ of the emitting element relative to the line of sight can be determined by $\cos\alpha=\cos\theta\cos\theta_{\rm v}+\sin\theta\sin\theta_{\rm v}\cos\phi$ \citep[see][for details]{Matsumoto2020}. In any case, what is directly relevant to our calculation is $L_{\rm iso}(\theta_{\rm v})$ rather than $E_{\theta}$.} in particular, if the large-angle emission is actually dominated by some scattering processes \citep{Kisaka2018}.

\begin{figure}
\centering\resizebox{0.45\textwidth}{!} {\includegraphics{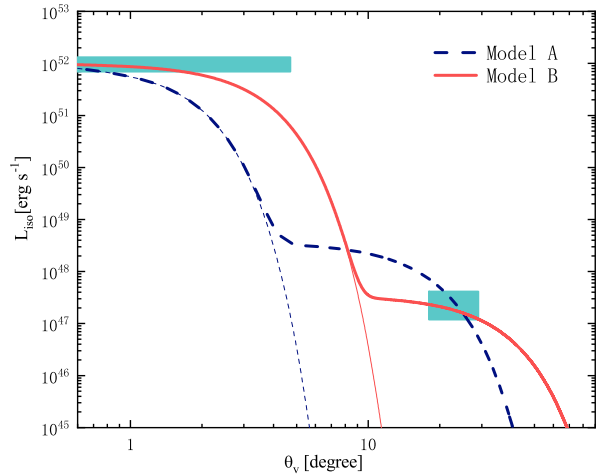}}
\caption{The isotropically-equivalent luminosity of SGRBs for different viewing angles. The rectangles give the luminosity ranges for the corresponding angles that are inferred from the GRB 170817A observations. The thin and thick lines depict the single- and two-Gaussian distributions, respectively. The parameter values for Model A and B are listed in Table 1, which are two representative best-fit models as discussed in Section 3.}
\label{fig2}
\end{figure}

As illustrated in Figure \ref{fig1}, the jet emission of a two-Gaussian distribution could have a more significant wing emission than the single-Gaussian case. Nevertheless, as long as the ratio between the two Gaussian components satisfies $\mathcal C\ll 1$, the SGRBs of a relatively high luminosity (e.g., the cosmological SGRBs) can still only be relevant to the inner Gaussian component. Therefore, when we fit the distributions of the fluxes and redshifts of the cosmological SGRBs in the next scetion, we will only take into account the inner Gaussian component. This means the outer Gaussian component is introduced and emphasized in this paper just in order to explain GRB 170817A simultaneously. However, this treatment is necessary and practicable for unifying GRB 170817A and the cosmological SGRBs, which is one thing we want to demonstrate in this paper. As further shown in Figure \ref{fig2}, the $L_{\rm iso}(\theta_{\rm v})$ function can first be constrained by the rectangles that are inferred from GRB 170817A. Different form Figure \ref{fig1}, the luminosity and the viewing angle in Figure \ref{fig2} are both plotted in logarithm scale. On the one hand, the observed isotropic luminosity of GRB 170817A is measured to $1.6^{+2.5}_{-0.4}\times 10^{47}\erg\s^{-1}$ for the viewing angle of $\theta_{\rm v}= 25^{+4}_{-7}$ degree. On the other hand, the on-axis luminosity of the jet of GRB 170817A for $\theta_{\rm v}\lesssim 5^{\circ}$ is found to be around $(1.0\pm0.3)\times 10^{52}\erg\s^{-1}$, which is required to explain the peak fluxes of the afterglow emission \citep[e.g., ][]{ Howell2019}.

In this paper, we term the probability distribution of the on-axis isotropic luminosity $L_{\rm on}\equiv L_{\rm iso}(0)$ as the intrinsic LF of SGRBs. The coupling of this intrinsic LF with a random distribution of the line of sights and as well as the telescope selections determines the luminosity distribution of the observed SGRBs. In many previous works, by assuming a top-hat jet structure, an empirical broken-power law LF can usually be derived from the observational luminosity distribution. 

Now, as a non-trivial jet structure is taken into account, it can be expected that the intrinsic LF of SGRBs could have a form simpler than the broken-power law. As an attempt, in the following calculations we assume the intrinsic LF to be a single power law with an exponential cutoff at the lower limit as
\begin{eqnarray}
\Phi(L_{\rm on})=\Phi_{*}\left({L_{\rm on}\over L_{\rm on}^{*}}\right)^{-\gamma}\exp\left(-{L_{\rm on}^{*}\over L_{\rm on}}\right).\label{SPL}
\end{eqnarray}
The availability of this empirical function will be judged by the final fittings to the observational distributions\footnote{Actually, we also test a broken-power law function and find that its low-luminosity segment increases very quickly, which makes it very close to the function presented in Equation \ref{SPL}. Then, following Occam's Razor, Equation (\ref{SPL}) is adopted.}
, as presented in the next section.


\begin{figure}
 \centering\resizebox{0.45\textwidth}{!}
	{\includegraphics{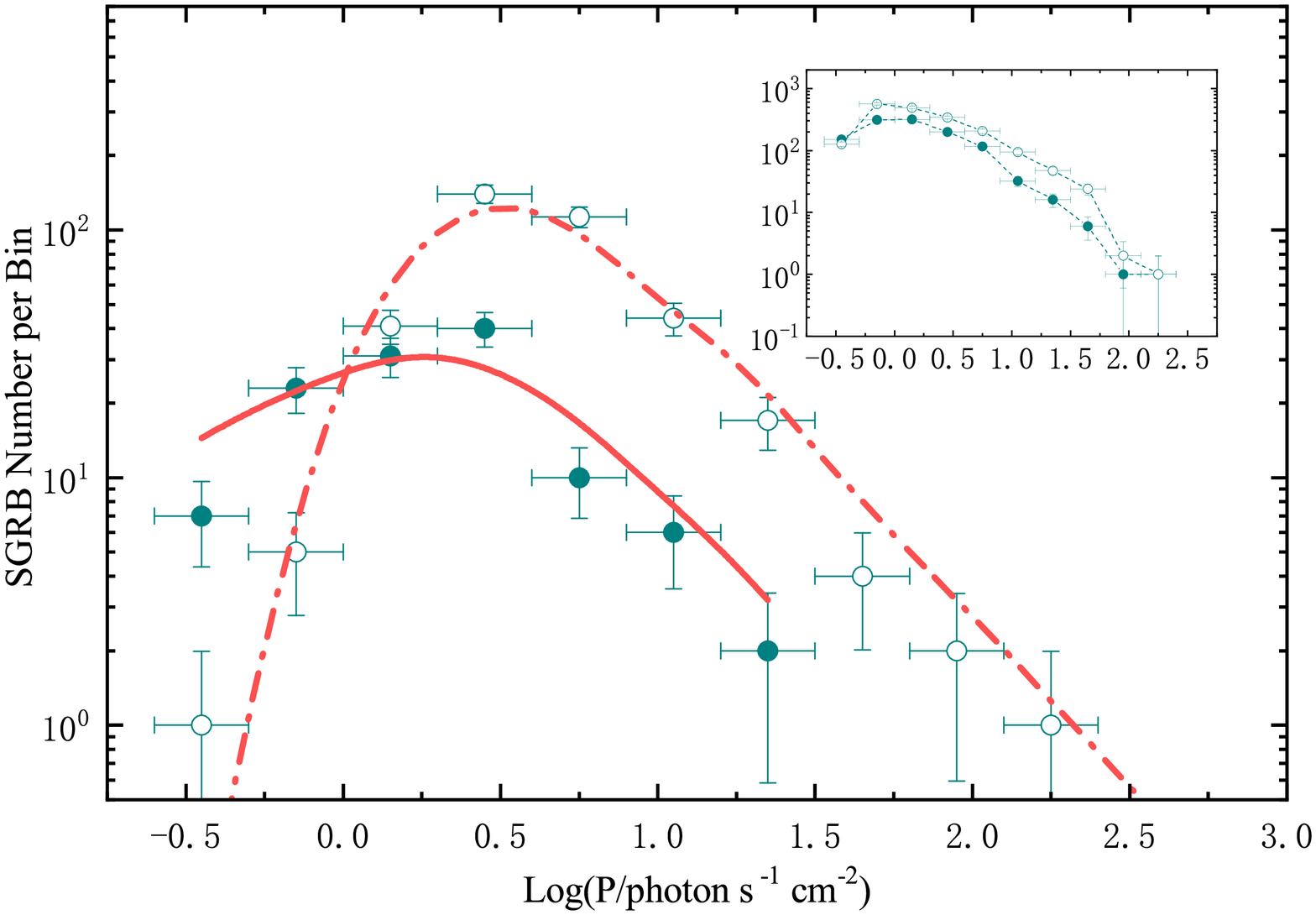}}
    \caption{The photon flux distributions of the cosmological SGRBs from the Fermi (open circles; in the energy band of $50-300\kev$) and Swift (solid circles; in the energy band of $15-150\kev$) observations. The lines provide the best fit to the data in Model B. The inset shows the flux distributions of the {\it Swift} and {\it Fermi} long GRBs. }
    \label{fig3}
\end{figure}
\begin{figure}
 \centering\resizebox{0.45\textwidth}{!}
	{\includegraphics{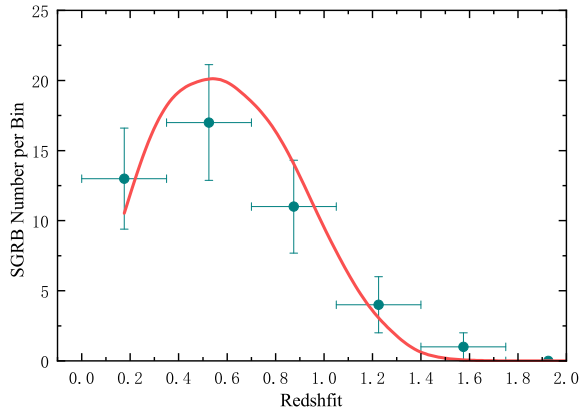}}
    \caption{The redshift distribution of the cosmological SGRBs (solid circles). The solid line gives the best fit in Model B.}
    \label{fig4}
\end{figure}

\section{Confronting the model with observations}

\subsection{The cosmological SGRB samples}

In the past 14 years, the Neil Gehrels Swift Observatory has detected more than 1300 GRBs including 119 SGRBs. The {\it Fermi} satellite began to hunt for GRBs from the year of 2008 and about 2300 GRBs have been captured, out of which 368 samples are SGRBs. Here, the SGRBs are defined as usual by a duration of $T_{90}<2\s$. In addition, the SGRBs having an extended emission are also included \citep[][]{Dietz2011, Kopac2012, Berger2014, Liu2019}. For the {\it Swift} GRBs, we take their $T_{100}$ photon fluxes in the energy band of $15-150 \kev$ from the web site of https:// swift.gsfc.nasa.gov/results/batgrbcat/index.html. These fluxes are obtained with the best-fit spectra of \cite{Lien2016} and averaged over the time interval during which $100\%$ of the burst emission is detected. Meanwhile, for the {\it Fermi} GRBs, their 64-ms peak photon fluxes fitted with a band function have been provided on https:// heasarc.gsfc.nasa.gov/W3Browse/fermi/fermigbrst.html \citep{Gruber2014,Kienlin2014,Bhat2016}, where two kinds of fluxes can be found in the energy bands of $50-300\kev$ and $10-1000\kev$, respectively. Specifically, the former kind of fluxes are adopted in this paper, since the energy band of $50-300\kev$ is the primary sensitive range of the {\it Fermi} Gamma-ray Burst Monitor (GBM). The distributions of the peak photon fluxes of the SGRBs are displayed in Figure \ref{fig3}. In comparison, the fraction of the low-flux samples (e.g., $P<1.4~ {\rm photons}\s^{-1}\cm^{-2}$) in the {\it Swift} SGRBs is obviously higher than that in the {\it Fermi} SGRBs. To a certain extent, this difference could be just due to the fact that the threshold flux of the {\it Fermi} GBM is somewhat higher than that of the {\it Swift} Burst Alert Telescope (BAT). However, it should still be noticed that the flux ranges of the long GRBs detected by {\it Fermi} and {\it Swift} are actually similar. By considering that the jet emission can decrease quickly with an increasing viewing angle, the low-flux SGRBs seems inclined to be detected off-axis. In view of the potential relevance between the luminosity and the hardness of the jet emission, the off-axis emission can be expected to be softer than the on-axis emission (e.g., Goldstein et al. 2017; Von Kienlin et al. 2019). Therefore, as a possible explanation, the difference in the flux distributions of the {\it Swift} and {\it Fermi} SGRBs could be somehow caused by the different ability of the BAT and GBM in detecting the soft off-axis emission at cosmological distances.

Additional to the flux distributions, an independent constraint on SGRB models can be provided by the redshift distribution. Nevertheless, in fact, it is not easy to measure a redshift for a SGRB, because of the usual poor localization, the extinction by the host galaxy \citep{Tanvir2008}. We collect the SGRB samples with a measured redshift from the GCN Circulars Archive on https://gcn.gsfc.nasa.gov/gcn3\_archive.html\#tc3 and literatures \citep{Dietz2011, Kopac2012, Berger2014, Liu2019}. There are totally 48 SGRBs detected by the {\it Swift} BAT, but most of them were missed by the Fermi GBM. The distribution of these redshift-measured SGRBs is shown in Figure \ref{fig4}.

\subsection{Theoretical distributions}
Because of the origin of SGRBs from the mergers of compact binaries, their event rate at redshift $z$ can in principle be connected with the cosmic star formation rates (CSFRs) by delay times due to the formation process of the
compact binaries and the orbital decay through gravitational radiation. Then, by considering that the delay time $\tau$ satisfies a probability distribution function $F(\tau)$, the SGRB rates can be calculated by \citep[e.g., ][]{Regimbau2009,Zhu2013,Regimbau2015}:
\begin{eqnarray}
&\dot{R}&_{\mathrm{SGRB}}(z)\propto(1+z)\int_{\tau_{\min}}^{t(z)-t(z_{\rm b})}{\dot{\rho}_{*}[t(z)-\tau]\over 1+z[t(z)-\tau]}F(\tau)d\tau\nonumber\\
&\propto&(1+z)\int^{z_{\rm b}}_{z[t(z)-\tau_{\min}]}{\dot{\rho}_{*}(z')\over 1+z'}F[t(z)-t(z')]{dt\over dz'}dz'\label{sgrbr},
\end{eqnarray}
where $\dot{\rho}_{*}(z)$ is the CSFR,  $t(z)=\int_z^{\infty}[(1+z')H(z')]^{-1}dz'$, $dt/dz=-[(1+z)H(z)]^{-1}$, and $z_{\rm b}$ represents the redshift at which the binaries started to be formed. The CSFR as a function of redshift can be empirically expressed by \citep{Robertson2012}
\begin{eqnarray}
\dot{\rho}_{\rm CSFR}(z)={a+b(z/c)^f\over 1+(z/c)^d}+g,
\end{eqnarray}
where $a=0.007 \msun \yr^{-1} \mpc^{-3}$, $b=0.27 \msun \yr^{-1} \mpc^{-3}$, $c=3.7$, $d=6.4$, $f=2.5$, and $g=3\times10^{-3}\msun \yr^{-1} \mpc^{-3}$.
A precise expression of the probability $F({\tau})$ is unaccessible because of the complexity of the binary evolutions and the orbital parameters, i.e., the initial separation and the initial ellipticity. Here, as usual, we take the empirical form as $F(\tau)\propto \tau^{-1}$ \citep{Piran1992,Guetta2006,Regimbau2009,Zhu2013,Regimbau2015},
which can be understood by the gravitational wave decays with power-law distributed orbital separations and a constant ellipticity.

For an SGRB of an observed flux $P$ in the energy band of $(E_1,E_2)$, its isotropic bolometric luminosity can be given by
\begin{eqnarray}
L_{\rm iso}=4\pi d_{l}^2Pk(z;E_{\rm p})\label{Liso1}
\end{eqnarray}
where $d_{l}$ is the luminosity distance. The $k-$correction factor is defined as
\begin{eqnarray}
k(z;E_{\rm p})\equiv{\int_{E_{\rm a}/(1+z)}^{E_{\rm b}/(1+z)}E S(E){\rm
d}E\over \int_{E_1}^{E_2}S(E){\rm d}E},
\end{eqnarray}
which converts the observational photon flux in the detector band ($E_1,E_2$) to the energy flux in a fixed rest-frame band ($E_a,E_b$), where $E_1=15\kev$ and $E_2=150\kev$ for {\it Swift}, $E_1=50\kev$ and $E_2=300\kev$ for {\it Fermi}, and as usual we take $E_{\rm a}=1$ keV and $E_{\rm b}=10^4$ keV. The energy spectrum $S(E)$ of the SGRB can be described by the Band function \citep{Band1993} with a spectral peak energy $E_{\rm p}$ and two spectral indices of typical values of $-0.5$ and $-2.3$ \citep[e.g.,][]{Yonetoku2014,Wanderman2015,Sakamoto2018}. The value of $E_{\rm p}$ can be determined by invoking an empirical $E_p-L_{\rm on}$ correlation as \citep{Tsutsui2013}
\begin{eqnarray}
L_{\rm on}=10^{52.29\pm0.066}\erg\s^{-1}\left[{E_{\rm p}(1+z)\over
774.5\kev}\right]^{1.59\pm0.11},\label{EPLP}
\end{eqnarray}
which is obtained by using a statistics of the cosmological SGRBs whose redshifts and peak energies were measured accurately\footnote{This empirical correlation is considered to be only available for the on-axis emission, since the SGRBs determining this correlation are all at redshifts $z\gtrsim 0.2$, where the wing emission is hard to be detected. Simultaneously, the highly off-axis emission of GRB 170817A is obviously inconsistent with this correlation.}. Simultaneously, the value of the on-axis luminosity $L_{\rm on}$ of the cosmological SGRBs can be given by
\begin{eqnarray}
L_{\rm on}\approx L_{\rm iso}\exp\left({\theta_{\rm v}^2\over 2\theta_{\rm in}^2}\right),\label{Liso2}
\end{eqnarray}
where the outer Gaussian component is ignored for these cosmological SGRBs. Finally, by combining Equations (\ref{Liso1}-\ref{Liso2}), we can get the expression of the isotropic luminosity of SGRBs as a function of their observed flux $P$ and viewing angle $\theta_{\rm v}$.

The differential detection probability of a SGRB is determined by the intrinsic LF and the viewing angle of the SGRB as
\begin{eqnarray}
dp=2\Phi(L_{\rm on}){2\pi\sin\theta_{\rm v}d\theta_{\rm v}\over 4\pi},
\end{eqnarray}
where the first number 2 represents the SGRB jets are paired.
Then, the detectable SGRB numbers in different flux ranges and different redshift ranges can be calculated by
\begin{eqnarray}
N(P_1, P_2)&=&{\Delta\Omega\over 4\pi} T \int^{z_{\max}}_{0}\int^{P_{2}}_{P_{1}}\int_0^{\theta_{\rm v, max}}\eta(P)\nonumber\\
&\times&\dot{R}_{\rm SGRB }(z) \Phi_{\rm}(L_{\rm on})\sin\theta_{\rm v}~d \theta_{\rm v} dP{dV(z)\over 1+z},
\label{EQN: PPFD}
\end{eqnarray}
and
\begin{eqnarray}
N(z_1, z_2)&=&{\Delta\Omega\over 4\pi} T \int^{z_2}_{z_1}\int^{P_{\max}}_{0}\int_0^{\theta_{\rm v,max}}\eta(P)\vartheta_z(z,P) \nonumber\\
& \times&\dot{R}_{\rm SGRB }(z)\Phi_{\rm }(L_{\rm on})\sin\theta_{\rm v}~d\theta_{\rm v}dP{dV(z)\over1+z}, \label{EQN: RD}
\end{eqnarray}
respectively, where $\Delta \Omega$ is the field of view of a telescope, $T$ is the working time with a duty cycle of $\sim$50\%, $\theta_{\rm v,\max}$ is the maximum viewing angle determined by the jet structure and the detector ability, $\eta(P)$ and $\vartheta(z,P)$ are the trigger efficiency and the probability of redshift measurement, respectively, and $dV(z)$ is the comoving cosmological volume element. The limit values of the redshift $z_{\max}$ and the flux $P_{\max}$ are taken according to the boundaries of the observational ranges.

The selection effects of the telescopes are very complicated. For the {\it Swift} BAT, its trigger probability as a function of the injected flux had been simulated by \cite{Lien2014}, according to which an empirical formula can be obtained as
\begin{equation}
{\eta (P)=}\left\{
\begin{array}{ll}
0   ,  &P< 5.5\times10^{-9} \erg\s^{-1}\cm^{-2},\\
{a(b+c P/P_{0})\over (1+P/d P_{0})} ,  &P \ge 5.5\times10^{-9} \erg\s^{-1}\cm^{-2},
\end{array}\right.\label{PTHF}
\end{equation}
where $a=0.47$, $b=-0.05$, $c=1.46$, $d=1.45$ and $P_{0}=0.6\times 10^{-7}\erg\s^{-1}\cm^{-2}$ \citep[e.g., ][]{Howell2014, Tan2015, Tan2018}. Meanwhile, however, such a simulation had never been carried out for the {\it Fermi} GBM and thus we have to adopt a cutoff threshold as usual as $P_{\rm th}=10^{-8} \erg\s^{-1}\cm^{-2}$. More strictly, the trigger probability is not only dependent on the injected flux, but also related to the energy of the injected photons, especially for the off-axis emission of a SGRB jet. For simplicity, we will include this effect into the adoption of the maximum viewing angle for the specific detectors. About the redshift measurement probability $\vartheta(P,z)$, we only pay attention to {\it Swift} since the redshift-measured SGRBs all belong to the {\it Swift} catalog. An empirical expression of $\vartheta(P,z)$ can be summarized from the statistics of the {\it Swift} long GRBs, which reads \citep{Cao2011, Coward2013, Tan2015}
\begin{equation}
\vartheta(P,z)=\zeta(P)\xi(z),
\end{equation}
where
\begin{equation}\label{zetp}
\zeta(P)={\rm Min}\left[0.27+\frac{P}{2\times10^{-6} \mathrm{erg}\ {\mathrm{s}}^{-1}\ {\mathrm{cm}}^{-2}},1\right]
\end{equation}
and
\begin{equation}\label{etz}
\xi(z)\propto \exp{\left(0.3-{z\over 8.9}\right)} \left\{1-0.41\exp\left[-{(z-1.6)^2\over0.11}\right]\right\}.
\end{equation}
Finally, please notice that the fluxes used in this paragraph are all in the unit of $\rm erg~s^{-1}cm^{-2}$, which is obtained by timing the peak photon fluxes to a coefficient of ${\int_{E_{\rm 1}}^{E_{\rm 2}}E S(E){\rm
d}E/ \int_{E_1}^{E_2}S(E){\rm d}E}$.

\subsection{Fittings and constraints}
By combining the distributions of the cosmological SGRBs and the observations of GRB 170817A , all model parameters can in principle be constrained. However, in fact, a very tight constraint is nearly impossible at present, because of the large number of the parameters and the high degeneracy between them. So, in this paper, we primarily care about a self-consistent explanation for the observations, but do not seek to a complete constraint on the model parameters. Then, for simplicity, the crucial parameters $\theta_{\rm in}$ and $\theta_{\rm out}$ will only be
assigned to a few of reference values and a relationship of $\theta_{\rm out}\sim10\theta_{\rm in}$ is taken by according to some simulation results \citep[e.g., ][]{Lazzati2017,Salafia2020}. The other model parameters will be constrained by the following two
separated steps:

\textit{(i) Fit the flux and redshift distributions of the cosmological SGRBs by fixing the parameter $\theta_{\rm in}$ and varying $\mathcal C$, $\gamma$, $\tau$, and $\dot{R}_{\rm SGRB}(0)$. Here only the inner Gaussian component is considered, as discussed in Section 2.}

\textit{(ii) Explain the event rate and the luminosity of GRB 170817A by fixing  $\theta_{\rm out}$ and varying ${\mathcal C}$, where the two Gaussian components are both involved and the inner Gaussian is described by the parameter values obtained from step one.}

\begin{table*}
\begin{center}
\caption{Constraints on the model parameters}\label{table1}
\begin{tabular}{c|c|c|c|c|c|c|c}
\hline\hline
\makecell*[c]Model&{$\theta_{\rm in}~ [^{\circ}]$}&$L_{\rm on}^{*}~[10^{52}\erg\s^{-1}]$&$\gamma$& $\tau_{\min}~[{\rm Gyr}]$ &\tabincell{c}{$\dot{R}_{\rm SGRB}(0)~\rm [\yr^{-1}\gpc^{-3}]$}& $\theta_{\rm out}~[^{\circ}]$  &  $\mathcal{C}[10^{-5}]$\\
\hline
\makecell*[c]A&1&$2.14^{+0.32}_{-0.32}$&$2.40^{+0.22}_{-0.19}$&$3.58^{+0.14}_{-0.14}$ &$1933.39^{+270.68}_{-116.00}$&10&$36.42^{+356.12}_{-31.75}$\\
\hline
\makecell*[c]B&2 &$2.85^{+0.39}_{-0.48}$&$2.42^{+0.25}_{-0.18}$&$3.53^{+0.13}_{-0.15}$ &$456.40^{+63.89}_{-27.38}$&20&$3.49^{+13.31}_{-2.11}$\\
\hline\hline
\end{tabular}
\end{center}
\end{table*}

\begin{figure}
\centering\resizebox{0.45\textwidth}{!} {\includegraphics{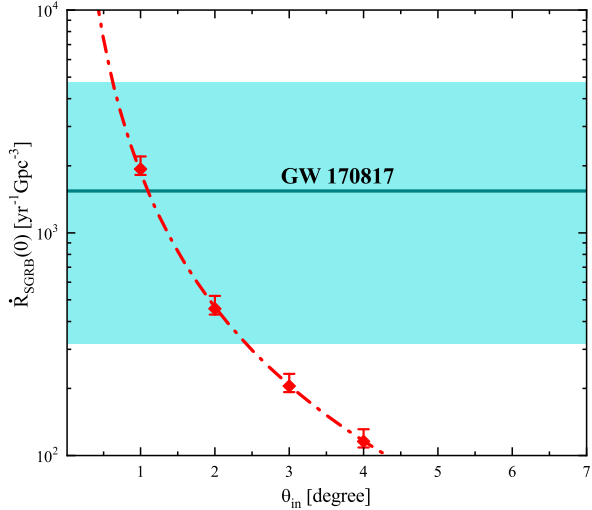}}
\caption{The dependence of the obtained value of $\dot{R}_{\rm SGRB}(0)$ on the choice of $\theta_{\rm in}$. The horizontal line represents the merger rate of $1540^{+3200}_{-1220}\gpc^{-3}\yr^{-1}$ inferred from the GW170817 event \citep{Abbott2017a}, while the shaded band represents the uncertainty.}
\label{fig5}
\end{figure}

\begin{figure}
\centering\resizebox{0.45\textwidth}{!} {\includegraphics{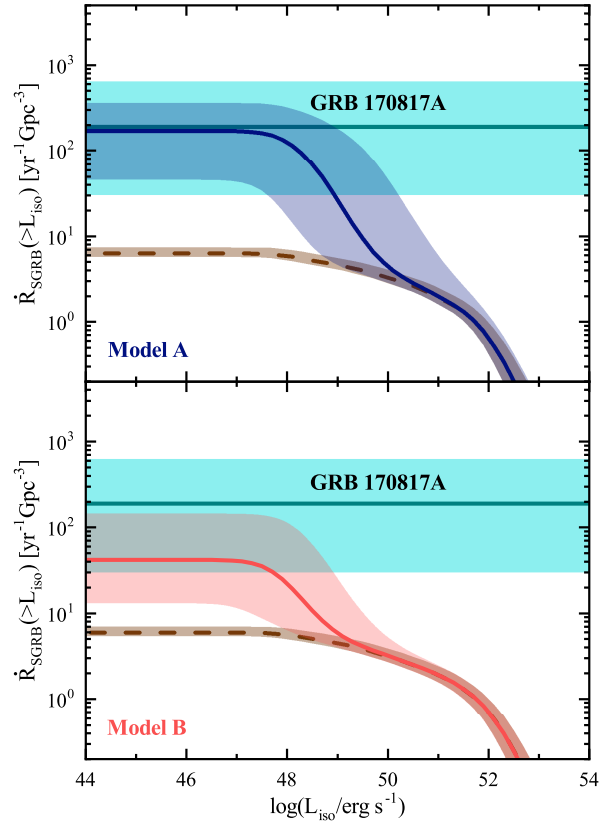}}
\caption{The local rate of SGRBs accumulated over the luminosities. The solid and dashed lines correspond to the cases with and without the outer Gaussian component, respectively. The horizontal line represents the event rate of $190^{+440}_{-160}\yr^{-1}\gpc^{-3}$ inferred from GRB 170817A. The shaded bands represent the uncertainties of the lines.}
\label{fig6}
\end{figure}

In step one, the goodness of the fits to the flux and redshift distributions is assessed by using the $\chi^2$ statistic test. The central values of the parameters can be obtained by minimizing $\chi^2$ and the 1-$\sigma$ errors are determined correspondingly, as listed in Table \ref{table1}. For different reference values of $\theta_{\rm in}$, we can obtain different constraints on the other model parameters. To be specific, the values of $L_{\rm on}^*$, $\gamma$, and $\tau$ are actually insensitive to the choice of $\theta_{\rm in}$. On the contrary, as shown in Figure \ref{fig5}, the uncertainty of $\theta_{\rm in}$ can lead the value of $\dot{R}_{\rm SGRB}(0)$ to vary in a wide range. Then, in comparison with the local rate of neutron star mergers of $1540^{+3200}_{-1220}\gpc^{-3}\yr^{-1}$ \citep{Abbott2017a}, which was inferred from the GW170817 event, we can conclude that the angle of the inner Gaussian $\theta_{\rm in}$ could not be much larger than $\sim3^\circ$. Otherwise, the constrained SGRB rate would be substantially lower than the merger rate and even lower than the rate of GRB 170817A. Therefore, in Table \ref{table1}, we only list two representative sets of parameters for $\theta_{\rm in}=1^\circ$ and $2^\circ$, which are termed as model A and B, respectively. Furthermore, in Figures \ref{fig3} and \ref{fig4}, we present an example fitting to the observational distributions of the cosmological SGRBs, with the central values of parameters in model B. Additionally, in these fittings, the upper limit of the integration over the viewing angle is required to be different for the {\it Swift} BAT and the {\it Fermi} GBM, i.e., $\theta_{\rm v,max}^{\rm Swift}\rightarrow{\pi\over 2}$ v.s. $\theta_{\rm v,max}^{\rm Fermi}\sim\theta_{\rm in}$. This can provide an effective explanation for the difference of the {\it Swift} and {\it Fermi} SGRBs in the low-flux distributions. It is indicated that the {\it Fermi} GBM could indeed be insensitive to the emission at $\theta_{\rm v}\gtrsim\theta_{\rm in}$, as suspected Section 3.1.

\begin{figure}
\centering\resizebox{0.45\textwidth}{!} {\includegraphics{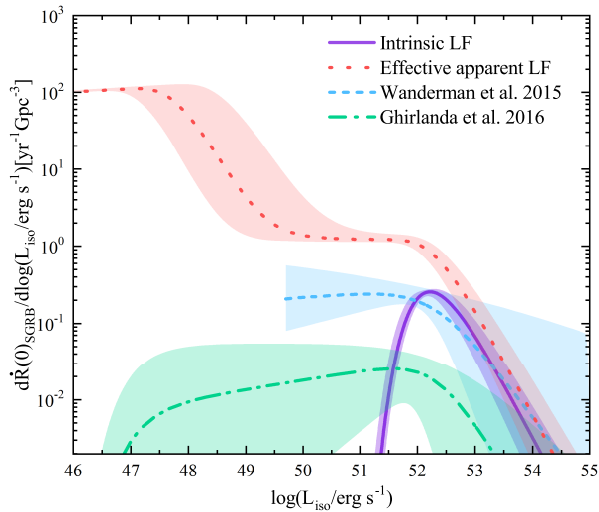}}
\caption{The apparent LF (dotted line) of SGRBs, which is obtained by combining the intrinsic LF (solid line) with the angular distribution of the jet emission. For a comparison, the previous apparent LFs discovered by \cite{Wanderman2015} and \cite{Ghirlanda2016} are also presented, the normalizations of which are shifted arbitrarily for a clear show. The shaded bands represent the uncertainties of the lines.}
\label{fig7}
\end{figure}

\begin{figure}
\centering\resizebox{0.45\textwidth}{!} {\includegraphics{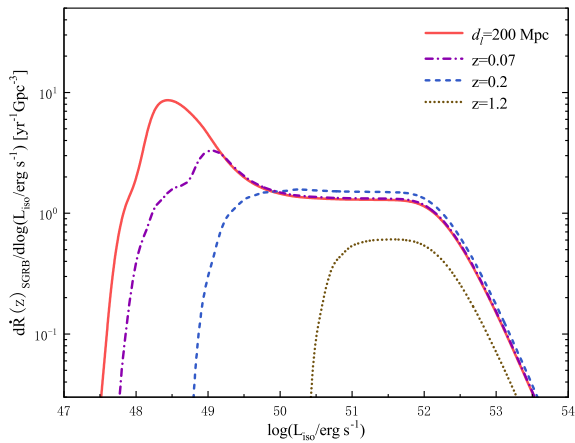}}
\caption{The predicted luminosity distributions of SGRBs at different distances for Model B parameters. A flux threshold of ${P_{\rm th}=10^{-8} \erg\s^{-1} \cm^{-2}}$ is used.}
\label{fig8}
\end{figure}

The fitting results obtained in step one clearly show that the single-Gaussian jet structure can be good enough to describe the cosmological SGRBs, as pre-assumed. However, what we want to emphasize here is that this single Gaussian model cannot be consistent with the observations of GRB 170817A. On the one hand, as shown by the thin lines in Figure \ref{fig2}, the required relatively low value of $\theta_{\rm in}\lesssim 3^\circ$ makes it difficult to explain the luminosity of GRB 170817A, because the wing emission of the single Gaussian decreases too quickly. Moreover, the really serious challenge to the single-Gaussian model is that its predicted local rate of detectable SGRBs for $L_{\rm iso}>10^{47}\rm erg~s^{-1}$ is drastically lower than the rate of GRB 170817A as $190^{+440}_{-160}\yr^{-1}\gpc^{-3}$ \citep{Zhang2018}. See the comparisons between the horizonal shaded band and the dashed lines in Figure \ref{fig6}. Therefore, it is natural to consider that the prompt emission of GRB 170817A is actually contributed by an extra outer-Gaussian component of the jet emission.

Then, in step two, we use GRB 170817A to constrain the parameters of the outer Gaussian. In view of the degeneracy between the parameters $\mathcal C$ and $\theta_{\rm out}$, we artificially take a fixed value for the outer Gaussian angle as $\theta_{\rm out}\sim10\theta_{\rm in}$, according to some simulation results \citep[e.g., ][]{Lazzati2017,Salafia2020}. Then, the value of $\mathcal C$ for GRB 170817A can be settled by accounting for its prompt luminosity as $1.6^{+2.5}_{-0.4}\times 10^{47}\erg\s^{-1}$ at the viewing angle of $\theta_{\rm v}= 25^{+4}_{-7}$ degree. The corresponding result is displayed by the thick lines in Figure \ref{fig2}. The obtained value of $\mathcal C$ is indeed very much smaller than 1. Here, this outer Gaussian component is suggested to be owned by all cosmological SGRBs and the value of $\mathcal C$ is considered to be universal. Then, as showed by the solid lines in Figure \ref{fig6}, the the predicted local rate of SGRBs of $L_{\rm iso}\gtrsim10^{47}\erg\s^{-1}$ can be effectively increased to be consistent with the rate of GRB 170817A, in particular, for relatively small $\theta_{\rm in}$.

To be summarized, the results presented in Figures \ref{fig2} and \ref{fig6} demonstrate that the cosmological SGRBs and GRB 170817A cannot be simultaneously explained by a single-Gaussian jet structure. Instead, a two-Gaussian structure can provide a very plausible explanation. Nevertheless, if only the cosmological SGRBs are concerned, then the inner Gaussian component alone could still be enough to account for their observational distributions (Figures \ref{fig3} and \ref{fig4}), since the outer Gaussian component is too weak to be detected at cosmological distances.

\section{The luminosity function and distributions}
In many previous works \citep[e.g., ][]{Wanderman2015,Ghirlanda2016}, the LF of SGRBs was usually determined from the observational luminosity distribution without considering of the angular distribution of the jet emission. As a result, a broken-power law LF was widely suggested, which however cannot directly reflect the physical distribution of the jet energies. Therefore, in this paper, we call such a LF as an apparent LF, in order to different from the intrinsic LF. With the two-Gaussian jet structure (Model B), we plot an apparent LF in Figure \ref{fig7} by integrating all observational directions. As shown, for relatively high luminosities ($\gtrsim10^{50}\rm erg~s^{-1}$), the apparent LF can be well described by a broken-power law, which is in good agreement with the previous discoveries \citep[e.g., ][]{Wanderman2015,Ghirlanda2016}. In other word, the previous apparent LFs can indeed be explained by the coupling of the intrinsic LF with the angular distribution of the jet emission. Specifically, the low-luminosity side of the apparent LF is completely contributed by the off-axis emission of SGRB jets. Furthermore, the contribution from the large-angle wing emission can only appear in the very low luminosity range.

Because of the telescope selection, the observational luminosity distributions of SGRBs are expected to evolve with the distance significantly, just as displayed in Figure \ref{fig8} for the Model B parameters and the GBM threshold. As shown, the low-luminosity fraction decreases rapidly with the increasing distance. The accumulated distribution presented in Figure \ref{fig6} is only available for a small distance, i.e., for $d_l\leq200$ Mpc. For $z\gtrsim0.1$, the contribution from the outer Gaussian disappears gradually and then it can be ignored safely, just as supposed when we fit the distributions of the cosmological SGRBs. On the one hand, for cosmological distances, if a detector is insensitive to the soft off-axis emission, then it can obtain a luminosity distribution very close to a single power law. On the other hand, for nearby SGRBs, it is possible to use their luminosity distribution to infer the angular distribution of their jet emission.

\section{Summary}
The observations of the GW170817/GRB 170817A event strongly indicate that SGRB jets have obvious angular structures, which make the observational luminosities of SGRBs sensitive to their viewing angles. Therefore, a certain fraction of the observed SGRBs, especially those of a relatively low luminosity, could actually be observed off-axis. The observational luminosity distribution can somewhat deviate from the intrinsic distribution of the jet energies. On the one hand, we revisit the fittings of the redshift and flux distributions of the cosmological SGRBs. It is found that these distributions can be well modeled with a single power-law intrinsic LF, while the jet off-axis emission is taken into account with an appropriate structure. As a result, the popular broken-power-law LF of SGRBs is demonstrated to be an observational manifestation of the combination of the intrinsic LF and the jet angular distribution. On the other hand, we further find that a two-Gaussian profile is an effective approximation for the jet structure, which is at least very helpful to self-consistently explain the event rate of GRB 170817A and the angular-dependence of its luminosity. Moreover, the inner Gaussian is constrained to be as narrow as $\theta_{\rm in}<3^\circ$, while the outer Gaussian is assumed to be about ten times wider than the inner one. Such a two-Gaussian jet structure could be a natural result of the propagation and breakout of the relativistic SGRB jets from a merger ejecta. The details of the afterglow emission from such a jet will be investigated in our following work.

With the parameters obtained in this paper, we can give some predictions for future GRB facilities. The Gravitational wave high-energy Electromagnetic Counterpart All-sky Monitor (GECAM) is a planed Chinese satellite, which is designed specially to monitor GW-associated SGRBs and will be launched in 2020. It has an all-sky field of view, a high sensitivity of $\sim 2\times 10^{-8}\erg\s^{-1}\cm^{-2}$, and a wide energy interval of $6\kev\sim 6 \mev$. By using these telescope parameters, we calculate the detectable SGRB numbers for GECAM, in particular, the numbers that can be associated by a GW signal. To be specific, in model A (model B), we have a number of $0.92^{+0.77}_{-0.53}$ ($0.44^{+0.45}_{-0.23}$), $6.96^{+5.87}_{-4.00}$ ($3.36^{+3.40}_{-1.75}$) and $21.06^{+19.02}_{-13.01}(9.39^{+11.04}_{-5.86})$ per year for a distance upper limit of $100$ Mpc, $200$ Mpc and $300$ Mpc, respectively. These distance limits are taken roughly corresponding to the aLIGO horizons in $O3$, $O4$ and $O5$, respectively, for neutron star mergers \citep{Abbott2018}.

\section*{Acknowledgements }
The authors appreciate the referee for his/her careful reading of the paper and valuable comments. This work is supported by the
National Natural Science Foundation of China (Grant Nos.
11803007, 11473008, 11822302 and 11833003), the Fundamental
Research Funds for the Central Universities (Grant
No. CCNU18ZDPY06), and the Science and technology research project of Hubei Provincial Department of Education (No. D20183002).

\end{document}